\input{aipcheck}

\documentclass[
    ,final            
  ]
  {aipproc}

\layoutstyle{6x9}

\begin{document}

\title{Spectral Analysis of Mid-IR Excesses of WDs}


\author{Jana Bil\'ikov\'a}{
  address={Astronomy Department, University of Illinois at Urbana-Champaign, 1002 West Green Street, Urbana, IL 61801, USA}
}

\author{You-Hua Chu}{
  address={Astronomy Department, University of Illinois at Urbana-Champaign, 1002 West Green Street, Urbana, IL 61801, USA}
}

\author{Kate  Y.-L. Su}{
  address={Steward Observatory, University of Arizona, 933 N. Cherry Ave., Tuscon, AZ 85721, USA}
}

\author{Robert A. Gruendl}{
  address={Astronomy Department, University of Illinois at Urbana-Champaign, 1002 West Green Street, Urbana, IL 61801, USA}
}

\author{Thomas Rauch}{
  address={Institute for Astronomy and Astrophysics, Kepler Center for Astro and Particle Physics, Eberhard~Karls~University, T\"ubingen, Germany}
}

\begin{abstract}
In our {\it Spitzer} 24 $\mu$m survey of hot white 
dwarfs (WDs) and archival {\it Spitzer} study of pre-WDs, 
i.e., central stars of planetary nebulae (CSPNs), we found mid-IR excesses for $\sim$15 WDs/pre-WDs. These 
mid-IR excesses are indicative of the presence of circumstellar dust that could be produced by sub-planetary 
objects. To further assess the nature of these IR-excesses, we have obtained {\it Spitzer} IRS, Gemini NIRI and 
Michelle, and KPNO 4m echelle spectra of these objects. In this paper we present the analysis of these 
spectroscopic observations and discuss the nature of these IR excesses.
\end{abstract}

\maketitle


\section{Dust Disk around the Central Star of the Helix Nebula}

The central star of the Helix Nebula is a hot white dwarf (WD) with an effective
temperature of $\sim110,000\,\mathrm{K}$. 
Its {\it Spitzer Space Telescope} MIPS 24 and 70 $\mu$m
observations have revealed a compact source coincident with the central WD.
A follow-up Infra-Red Spectrograph (IRS) observation of the central point source
has confirmed that the mid-IR emission originates from a dust continuum
with a temperature of 90$-$130 K, and an emitting area of 4$-$40 AU$^2$.
Only an extended object, such as a dust disk, can explain these properties.
The location of the dust, 40$-$100 AU, corresponds to the location of the Kuiper Belt
in our Solar System, and the dust disk was suggested to originate from collisionally
disrupted Kuiper Belt-like objects (KBOs) dynamically rejuvenated in the AGB and
post-AGB evolutionary stages \citep{Su07}.

\section{Spitzer MIPS 24 $\mu$m Survey of Hot WDs}

To search for more dust disks similar to that around the WD in the Helix Nebula,
we have carried out a {\it Spitzer} MIPS 24 $\mu$m survey of 71 hot ($\sim100,000\,\mathrm{K}$) WDs.
A compact 24 $\mu$m source coincident with the WD is detected in 9 cases;
in 7 of these, the star is still surrounded by a PN (Chu et al., 2010, in preparation).
We have constructed the SEDs of these WDs using optical and near-IR photometry
from the literature. All detections show excess emission at 24 $\mu$m. 
For four of these WDs, we have acquired follow-up {\it Spitzer} IRS spectra,
and all show dust continuum emission, with some showing additional emission lines.
The images and SEDs of three of these targets are shown in Figure 1, 
and described below.

{\bf Sh 2-216}, at a distance of 219 pc \citep{Harris07}, is the closest PN.
The SED of its central star follows the blackbody curve from optical to IRAC bands, but shows large
excess at 24 $\mu$m. The CSPN is not detected at 70 $\mu$m; and the SED shows the  
 3-$\sigma$ upper limit. The non-detection at 70 $\mu$m places constraints on the outer disk radius.
The {\it Spitzer} IRS spectrum is shown in the SED as gray symbols, and the smoothed spectrum is displayed as a thick line.
The spectrum is dominated by featureless continuum emission, which starts rising at $\sim$10 $\mu$m.
Such SED is similar to that of the Helix nebula's CSPN.

We model the dust emission using an optically thin dust model, assuming that the dust is heated by the central WD.
The spectrum does not show mineralogical features; we assume the composition to be astronomical silicates.
For the radiation from the WD, we adopt the synthetic spectral model from \citet{Rauch07},
the associated WD effective temperature of 95,000 K and log g of 6.9.

Small grains will be blown out of the system due to radiation pressure, and the minimum grain size can
be estimated using the ratio of radiation pressure force to gravitational force ($\beta$), which depends on
WD's luminosity and mass, and the dust grain density. We assume that the
grains will be blown out for $\beta$ of 0.5 \citep{Artymowicz97}.
To calculate the luminosity of the WD, we use the distance of 129 pc \citep{Harris07}, and
integrate the synthetic spectrum normalized to fit the optical and near-IR photometry. We find the WD
luminosity of $\sim$40 $L_{\odot}$. Note that this value is a lower than that derived by
\citet{Rauch07}, 158.5 $L_{\odot}$, because of their larger spectroscopic distance, 224 pc.
Using the WD luminosity of 40 $L_{\odot}$, a mass of
0.55 $M_{\odot}$ \citep{Rauch07}, and a dust grain density of 2.5 g cm$^{-3}$,
we find the minimum grain size to be $\sim$35 $\mu$m. We use a maximum grain size of 1 mm, and
assume that larger grains will not contribute significantly to the IR emission.
The absorption coefficients are calculated using Mie theory.
We also assume a power-law grain size distribution with a power index of -3.5, i.e., n(a)$\propto$a$^{-3.5}$,
typical of collisionally produced dust, and a uniform disk surface density.

The observed fluxes, IRS spectrum, and the 70 $\mu$m non-detection can be approximated by 
emission from a dust disk at radii  $\sim$50 to 80 AU, and a mass of $\sim$0.001 $M_{\oplus}$.
Note that the uncertainty in distance affects the calculation of the WD luminosity, which sets the
minimum grain size, and subsequently affects the disk's physical parameters.
Further exploration of the model's uncertainties is needed.

The {\it HST} observations of {\bf CSPN  K1-22} resolved a red companion 0.35$''$ away from the CSPN \citep{Ciardullo99}.
The SED in Figure 2 shows $V$ and $I$ magnitudes from each star individually \citep{Ciardullo99}, the remaining magnitudes are 
for the two stars combined. The CSPN emission is approximated by 
a blackbody, and the emission from the companion is a Kurucz model for an M0V star.
The two solid curves show contributions from the two components, and the dashed curve shows the 
sum of these two components.
The IR flux densities are all higher than the expected photospheric emission of these two stars.
The IRS spectrum shows a continuum component, as well as a strong [OIV] 25.89 $\mu$m emission
line component, both of which may be contributing to the observed 24 $\mu$m excess, but 
the line emission may also be 
attributed to incomplete local background subtraction.
Due to a poor angular resolution, it is unclear whether the 24 $\mu$m source is centered 
on the CSPN or its red companion, and future high-resolution mid-IR imaging is necessary 
to examine the coincidence of the 24$\mu$m source and the CSPN.

For purposes of dust disk modeling, we assume that the dust surrounds the CSPN,
and the companion is too far and too cool to contribute significantly to the heating of the dust.
We approximate the radiation from the CSPN by a blackbody with an effective temperature of
141,000$\pm$31,000 K \citep{Rauch99}, at a distance of 1.33 kpc \citep{Ciardullo99},
normalized to fit the observed optical fluxes. Such approximation yields a luminosity of
325 L$_{\odot}$, which, together with the mass of 0.59 $M_{\odot}$ \citep{Rauch99},
suggests a minimum grain size of $\sim$250 $\mu$m.
The observed dust continuum can be approximated by a disk extending from 
the sublimation radius, $\sim$0.62 AU, to $\sim$40 AU,
with a dust mass of $\sim$0.002 $M_{\oplus}$.

Note that the distance to the CSPN is very uncertain, and
different authors report values between 1.33 and 3.43 kpc.
The large uncertainty in distance introduces a large uncertainty to our dust models,
since the distance affects the luminosity calculation, which in turn
affects the minimum grain size, and the physical properties of the dust disk.

Another dust disk candidate with SED similar to that of the Helix CSPN is the central star of
EGB\,1, {\bf WD0103+732}.  {\it HST} images do not show any companion stars \citep{Ciardullo99}. 
The SED shows optical and near-IR 
flux densities following the blackbody curve of the hot central WD. 
2MASS $H$ and $K$ data points show upper limits.
Follow-up IRAC observations from 
{\it Spitzer's} Cycle 5 show that IRAC 3.6, 4.5 and 5.8 fluxes also lie on the blackbody tail,
the flux at 8 $\mu$m is above the photospheric emission level, and the MIPS 24 $\mu$m band flux 
density is more than three orders of magnitude higher than the expected photospheric emission.

The 24$\mu$m image shows that the WD is superposed on diffuse emission, which is dominated
by line emission. The background-subtracted spectrum is dominated by dust continuum emission.
To model the emission from the WD itself, we use the $UBVRIJHK$ photometry from literature and the 2MASS catalog,
and the WD parameters from \citet{Napiwotzki01}: distance of 650 pc, 
effective temperature of 147,000 K, and stellar mass of 0.65 $M_{\odot}$.
Using the blackbody approximation for the WD
normalized to fit the optical and near-IR fluxes, we get a luminosity of
$\sim$480 $L_{\odot}$.
Therefore, dust grains smaller than $\sim$340 $\mu$m will be blown out of the system.
Preliminary modeling suggests a dust disk
between $\sim$200 and $\sim$360 AU, and a dust mass of $\sim$0.14 $M_{\odot}$.

\begin{figure}[!t]
\resizebox{0.38\columnwidth}{!}
 {\includegraphics{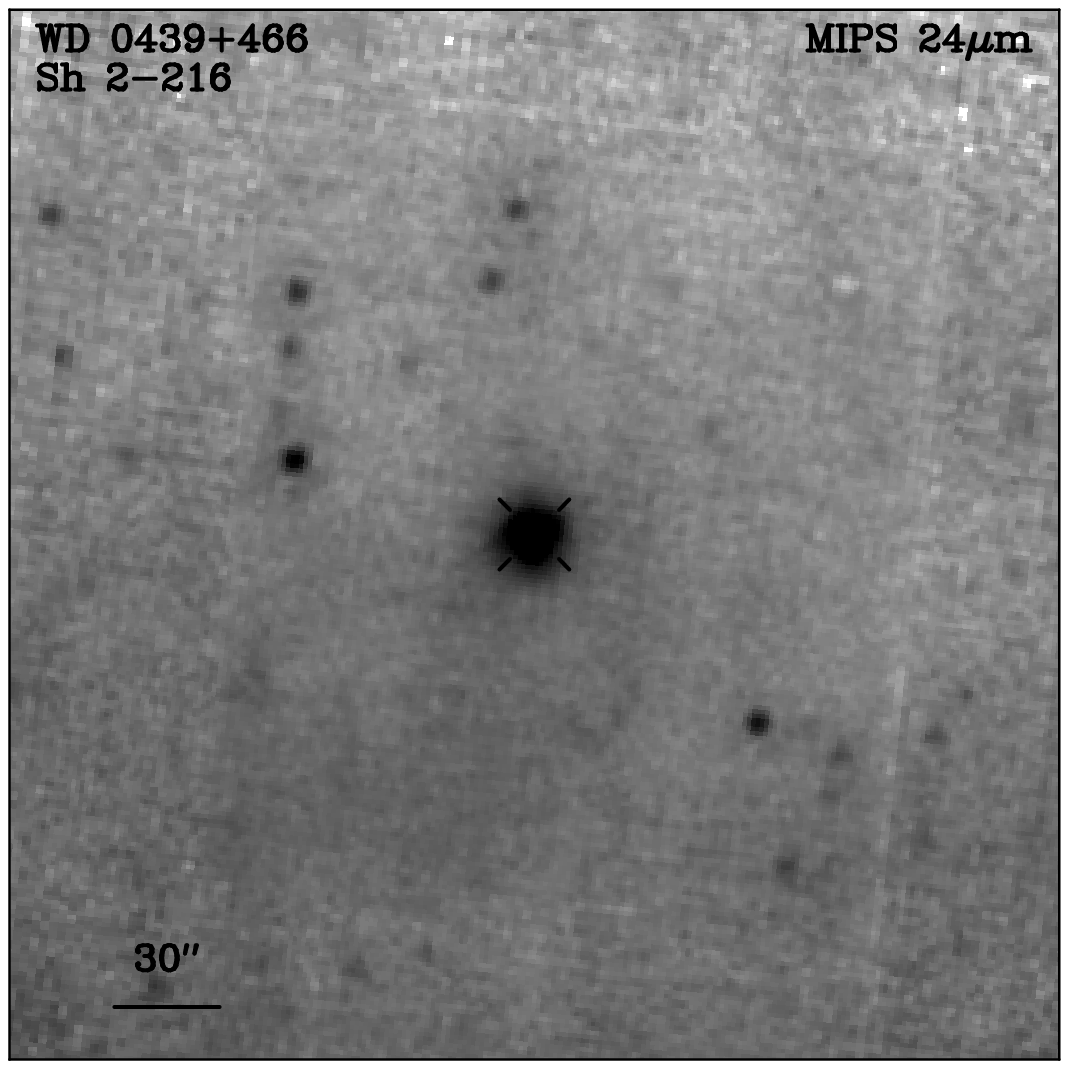}}
 \resizebox{0.52\columnwidth}{!}
 {\includegraphics{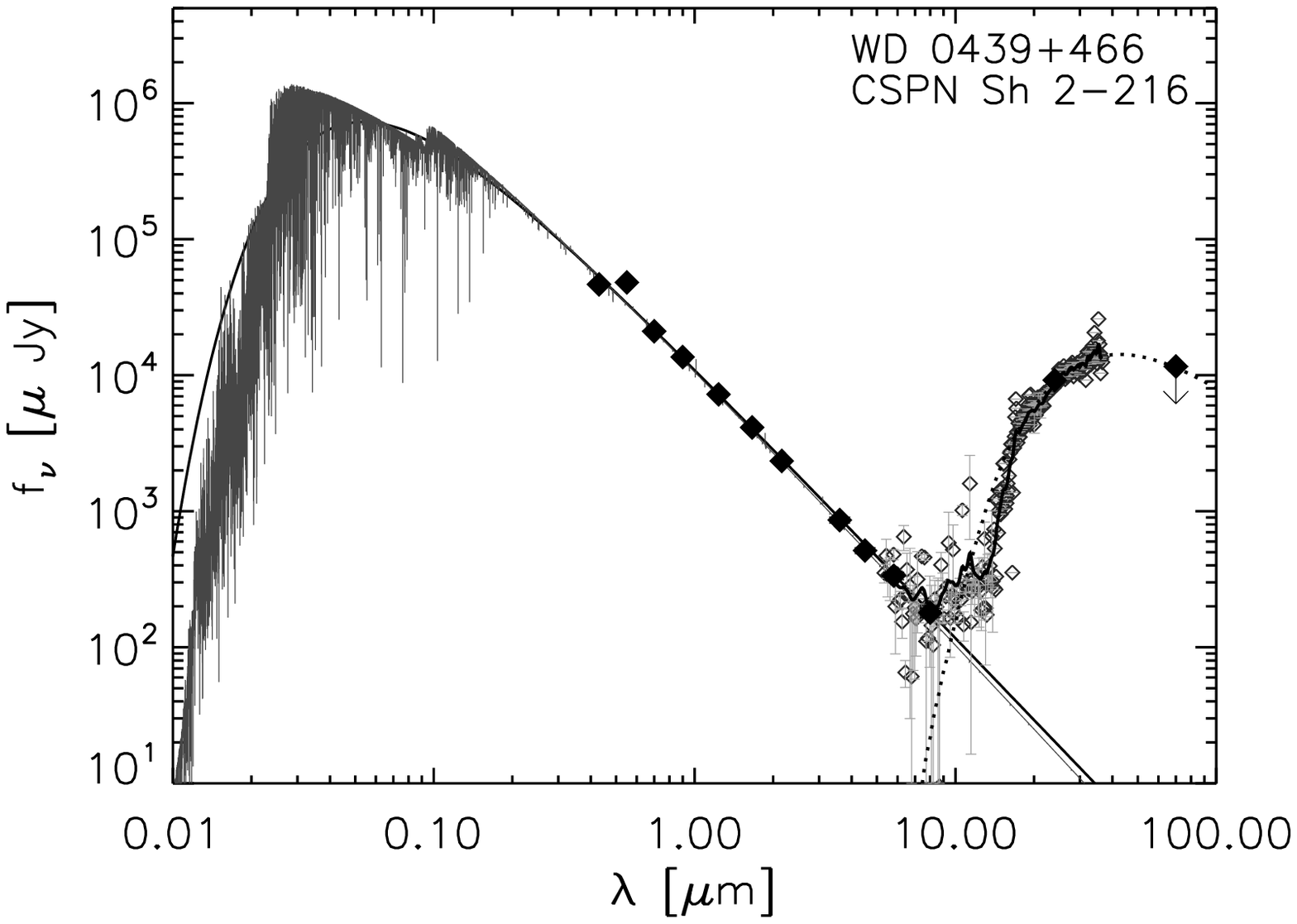}}
\end{figure}
\begin{figure}[!t]
\resizebox{0.38\columnwidth}{!}
 {\includegraphics{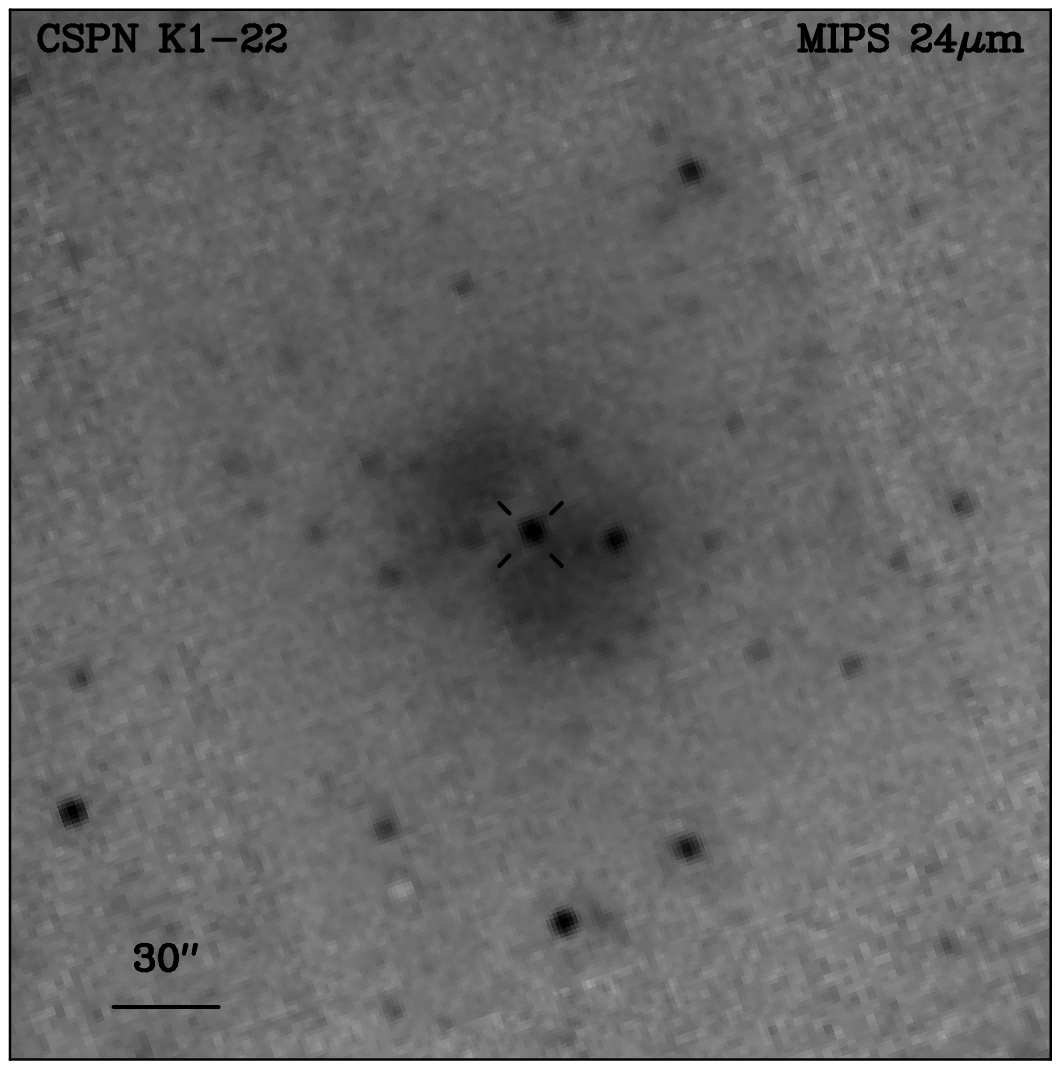}}
 \resizebox{0.52\columnwidth}{!}
 {\includegraphics{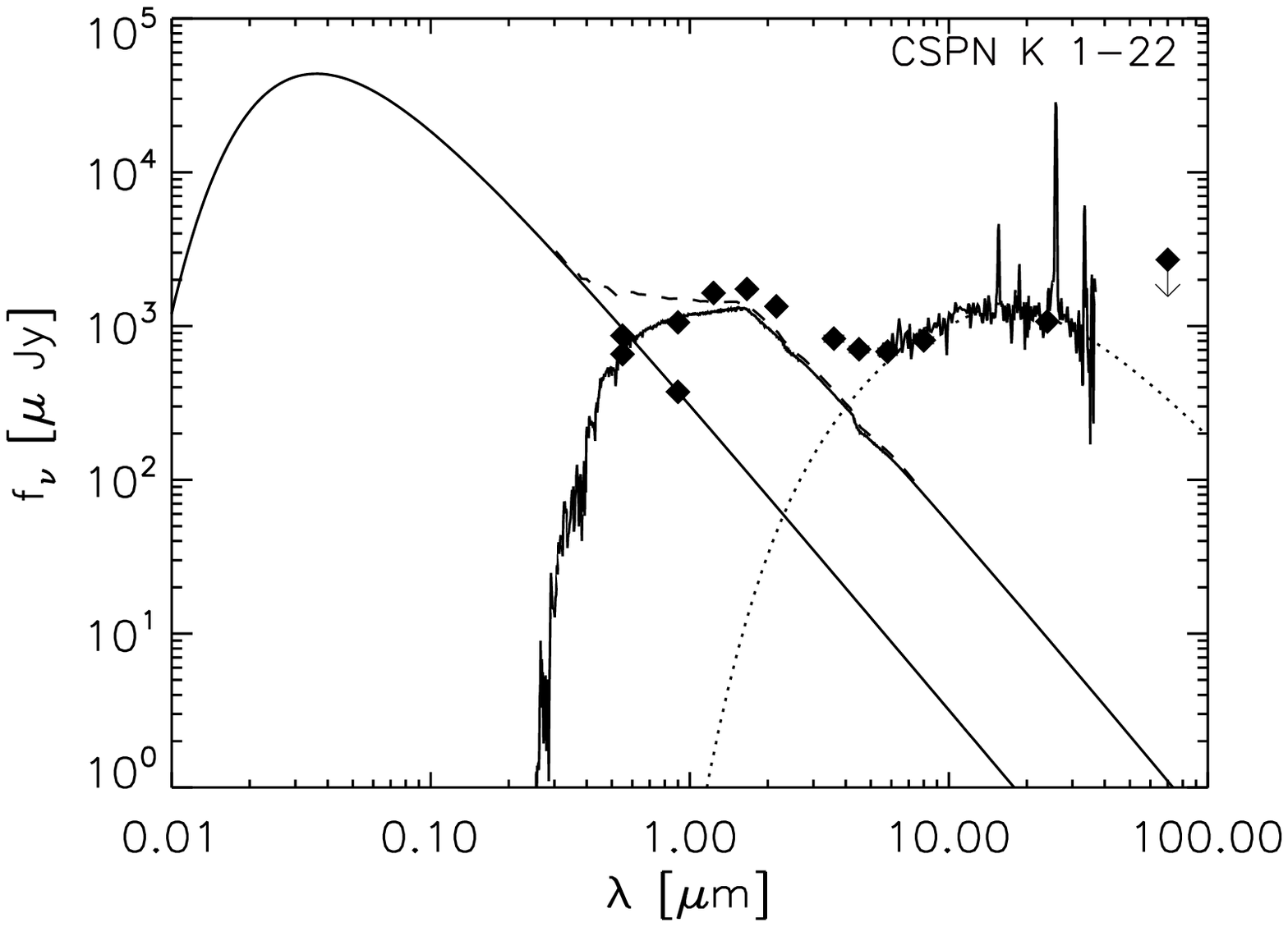}}
\end{figure}
\begin{figure}[!t]
\resizebox{0.38\columnwidth}{!}
 {\includegraphics{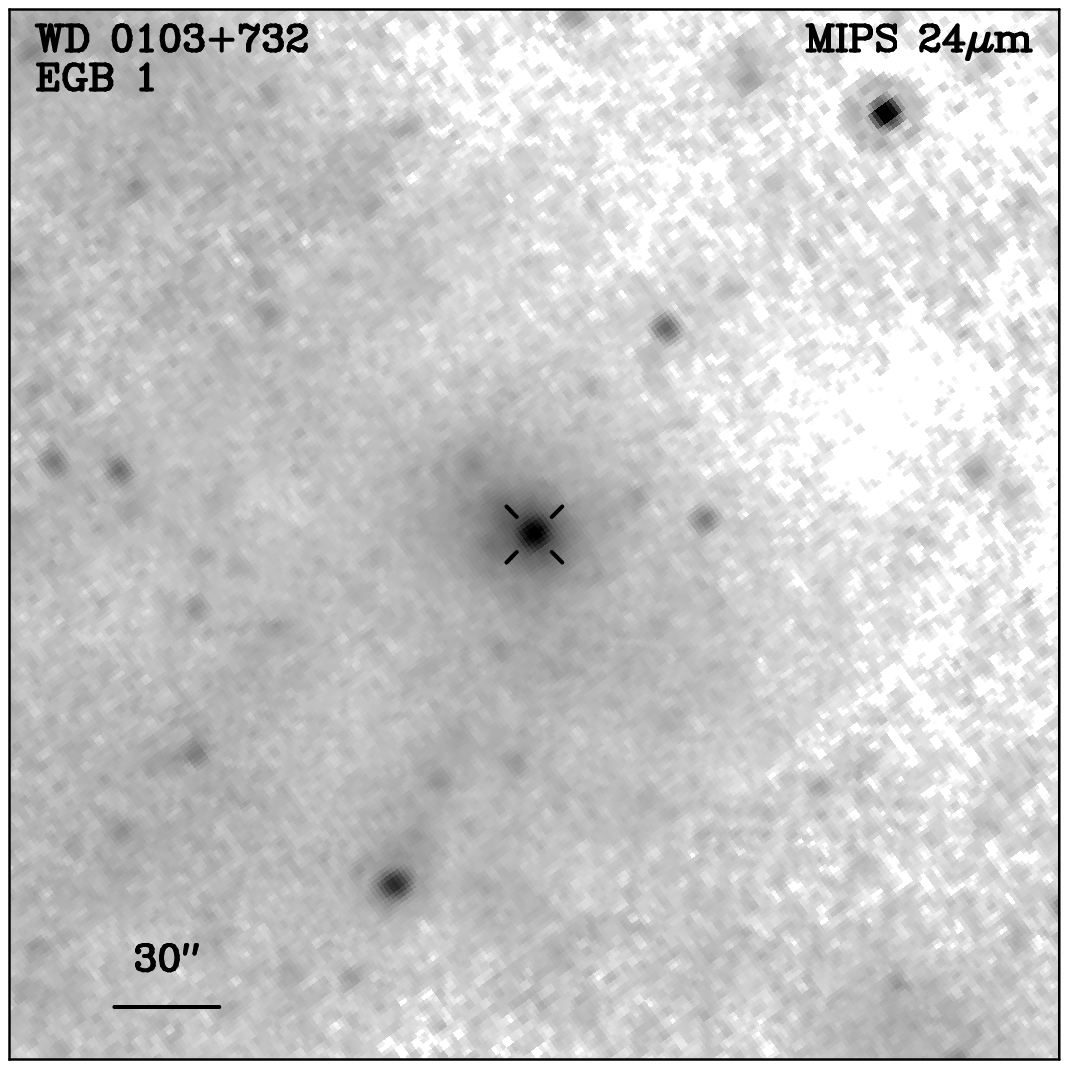}}
\resizebox{0.52\columnwidth}{!}
 {\includegraphics{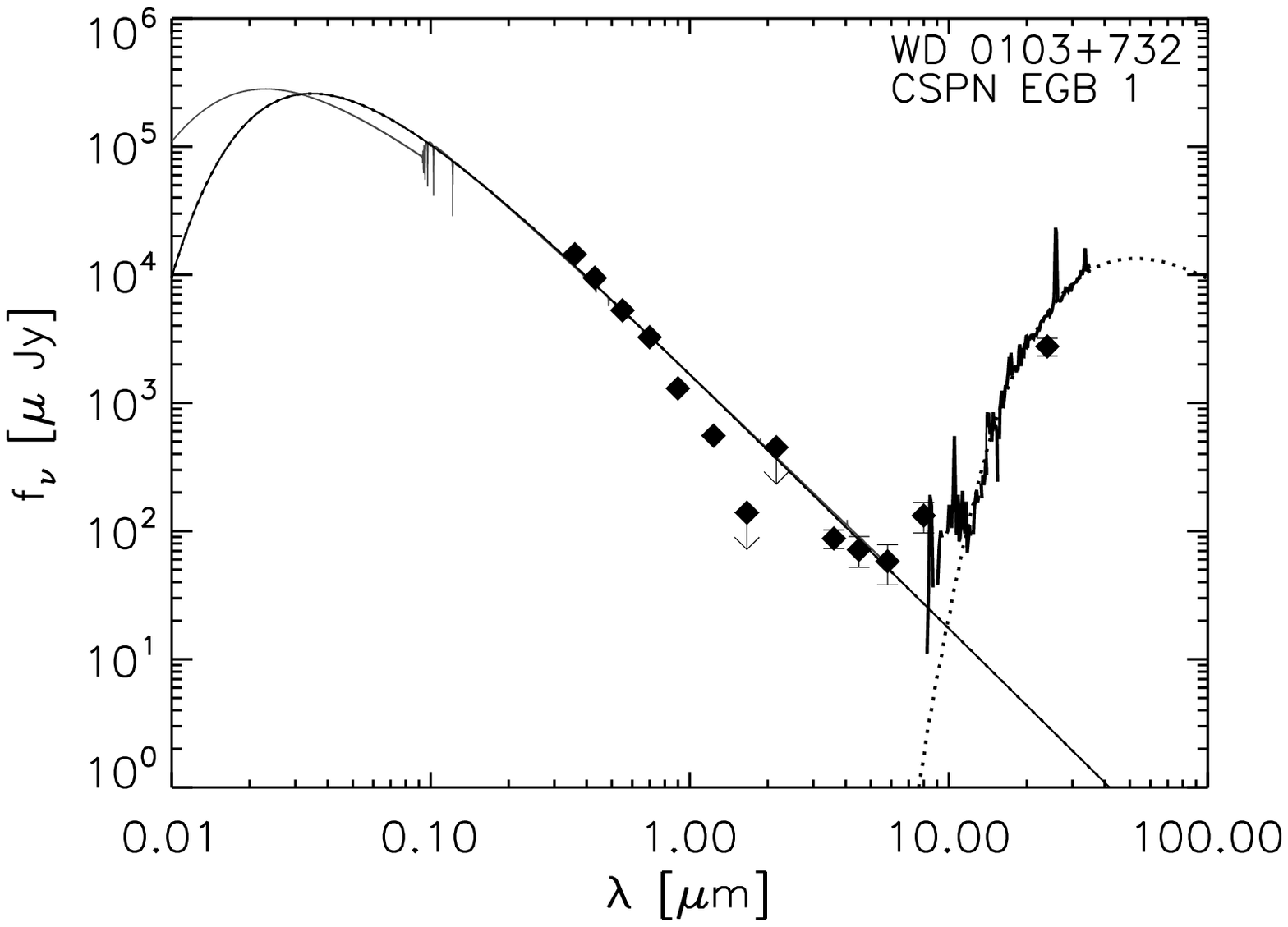}}
\caption{\footnotesize SEDs of Sh 2-216, K1-22, and EGB 1 from {\it Spitzer} 24$\mu$m survey.
The SEDs are constructed with optical photometry from literature, 2MASS $JHK$, 
IRAC photometry and {\it Spitzer} MIPS 24 $\mu$m data.
The IR spectra from {\it Spitzer} IRS are shown in solid line for K1-22 and EGB 1.
For CSPN Sh2-216, the IRS spectrum is shown in gray symbols, and the smoothed
spectrum is shown in solid line.
The dotted line represents the dust disk model.}
\label{fig:1}
\end{figure}

\section{Spitzer Archival survey of CSPNs} 
Since most cases of hot WDs exhibiting 24 $\mu$m excesses are still surrounded by PNe,
we have used archival {\it Spitzer} IRAC
(3.6, 4.5, 5.8, and 8.0 $\mu$m) and MIPS (24, 70, and 160 $\mu$m)
observations of PNe to search for CSPNs with IR excesses.
We have examined images of 66 resolved PNe, and selected 18 in which
the nebular emission was not too confusing or dominant in the central region,
and the CSPN was detected in most IRAC bands.
For these 18 cases, we have carried out photometric measurements, and constructed
the SEDs. Six of these CSPNs show convincing IR excesses.

In the case of NGC 6804 (Figure 2), IR excess is seen 
starting from $J$ band throughout all IRAC channels. 
The SED does not exclude the possible presence of a
cool ($T_\mathrm{eff}\approx 1500\,\mathrm{K}$) companion; however, a companion
alone cannot account for all of the observed IR excess, and
no companion has been detected around this CSPN \citep{Ciardullo99}.
Figure 2 presents a follow-up {\it Spitzer} MIPS 24$\mu$m image,
which shows a central source coincident with the CSPN.
The SED displays a follow-up Gemini NIRI 1--5 $\mu$m spectrum,
which reveals a rising continuum, as well as a compact emission line source.
Furthermore, Gemini Michelle 8--15 $\mu$m spectrum
exhibits a 10 $\mu$m silicate emission feature.

\begin{figure}
\includegraphics[width=0.39\textwidth]{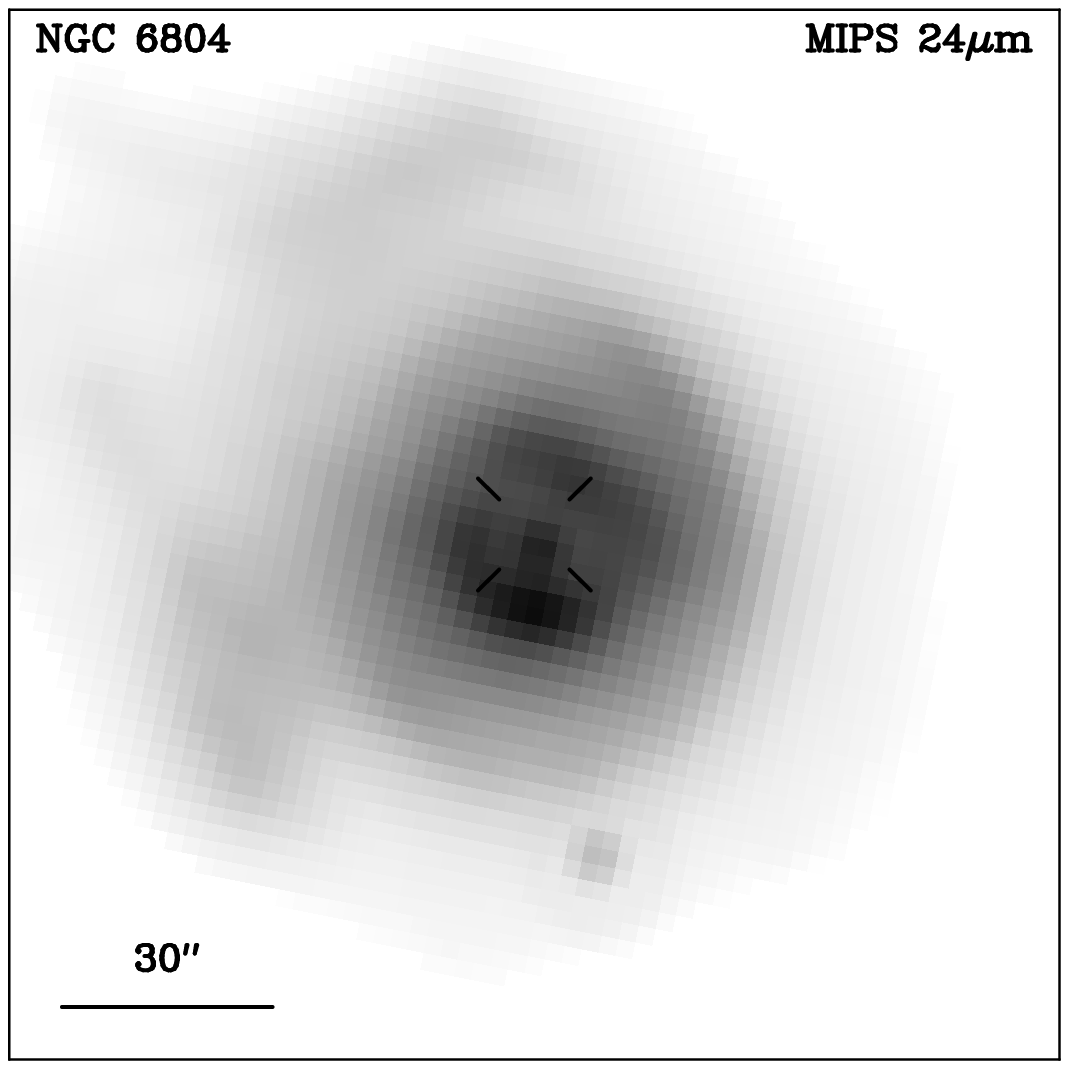} 
\includegraphics[width=0.53\textwidth]{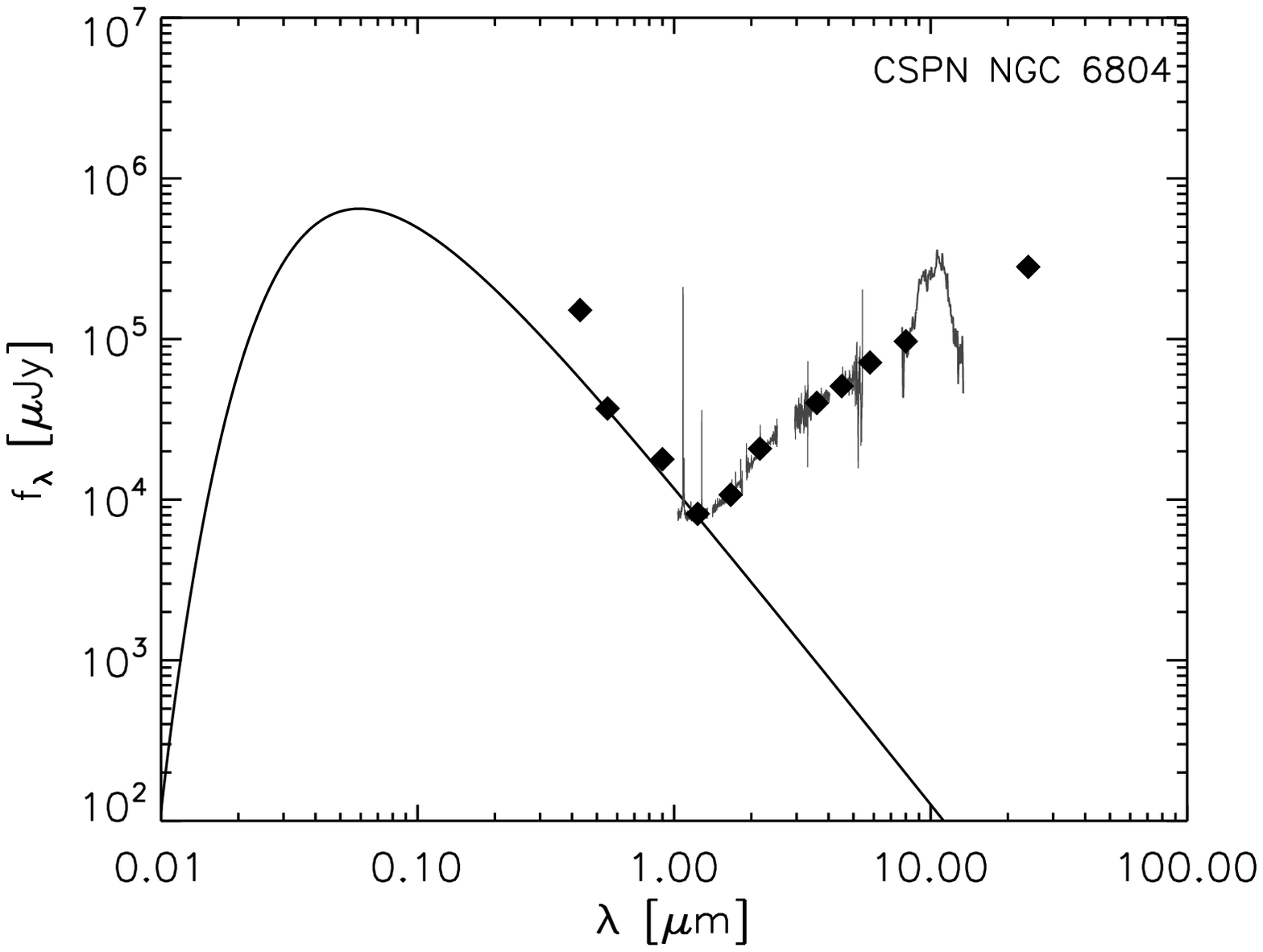}
\caption{{\it Spitzer} 24 $\mu$m image (left) and SED (right) of 
CSPN NGC 6804.
The SED is constructed with optical photometry from literature, 2MASS $JHK$, 
IRAC photometry and {\it Spitzer} MIPS 24 $\mu$m data.
The IR spectra from Gemini's NIRI and Michelle are shown in grey.
 }
\end{figure}

\section{Possible Origins of IR excesses}
The differences between the SEDs of our targets 
imply different properties of the disks, if not even different origins of
excess emission. We need to consider origins other
than the collisional disruption of KBOs that can 
produce the observed IR excesses.

\subsection{Post-AGB Binary Evolution}
It has been suggested from an analysis of broad-band
SEDs of 51 post-AGB stars
that Keplerian rotating dust disks are common among
binary post-AGB stars \citep{deRuyter06}. \citet{vanWinckel03}
showed that post-AGB stars displaying SEDs of warm dusty disks are
all single-lined spectroscopic binaries and the hot dust persists in the system 
because it is trapped in a stable circumstellar or circumbinary orbit.

Seven out of 9 hot WDs with 24 $\mu$m excesses from our survey
and the CSPNs with IRAC and/or MIPS excesses found in the {\it Spitzer} archive are
in PNe and thus represent the youngest WDs that have just
evolved past the post-AGB phase. One of them (NGC\,2346)
has a confirmed binary companion \citep{Mendez81} and SED that resembles
those of post-AGB binaries; thus, one cannot help asking 
whether some of these 24 $\mu$m and/or IRAC excesses are also related to the 
IR excesses of binary post-AGB stars reported by \citet{deRuyter06}.

However, it is difficult to identify and confirm the presence of a close companion
of a CSPN via direct imaging \citep{Ciardullo99}, and irregular spectral
variations due to winds hamper the detection of periodic radial velocity
variations \citep{DeMarco07}.
If a dust disk trapped in a stable orbit around a binary system persists throughout the PN phase, its
presence can serve as a powerful diagnostic for the binarity of a CSPN.
Further observational and theoretical studies are necessary to
distinguish between the two origins.

\subsection{Compact Unresolved Nebulosity}

Another possibility is that the observed excess comes from a compact 
nebulosity with high dust-to-gas ratio, which is seen in the born-again PNe 
Abell 30 and Abell 78 \citep{Cohen74}.
This central emission enhancement originates from knots seen in [OIII] $\lambda$5007 \AA \, and He II $\lambda$4686 \AA ,
but undetected in H$\alpha$ \citep{Jacoby79, Hazard80}.
Such observed morphological difference implies H depletion in
the central region \citep{Jacoby83}, which can occur in born-again PNe \citep{Iben83}.
We have carried out KPNO echelle spectroscopy of the hot WDs with 24 $\mu$m excesses
and CSPNs with IRAC excesses, covering both H$\alpha$ and [OIII] $\lambda$5007 \AA\, lines.
In all observed cases, each [OIII] feature has its H$\alpha$ counterpart. 
Therefore, born-again scenario cannot explain the observed IR excesses of most of our targets.

%

\section{Summary}

The discovery of a dust disk around the CSPN of the Helix nebula through its 24 $\mu$m excess
has inspired us to conduct a {\it Spitzer} 24 $\mu$m survey of hot
($T_\mathrm{eff} \approx 100,000\,\mathrm{K}$) WDs.
Out of 71 targets observed, 9 show 24 $\mu$m excesses, 7 of them in PNe.
To find more cases of CSPNs with IR excess, we have searched the {\it Spitzer} archive,
and found 6 targets with convincing IR excesses.
While some SEDs are similar to that of the Helix CSPN and show excess emission only
in mid-IR, others show excess starting at shorter wavelength.
Similarly, while some mid-IR spectra are dominated by dust continuum, others show
strong emission lines superposed on dust continuum.
The dust around the Helix central star was suggested to be produced by collisions among KBOs \citep{Su07}.
However, other mechanisms that could produce IR excess, such as binary interactions, need to be considered as well.
The SEDs and models show a variety of properties. Careful spectral modeling
and characterization of the central star are needed to accurately determine the disk properties.
Future modeling of the mid-IR SEDs is needed to evaluate the origins of the observed
excess emission.

\bibliographystyle{aipprocl} 



\end{document}